\documentclass{article}
\usepackage{spie,psfig,amsmath,amssymb}

\title{Abundances of Stars in the Galactic Bulge Obtained Using the
Keck Telescope\supit{1}}
\author{R. Michael Rich\supit{a} and  A. McWilliam\supit{b}  \skiplinehalf
  \supit{a}University of California, Los Angeles, Division of\\
  Astronomy and Astrophysics, 405 Hilgard Avenue, Los Angeles, CA
  90095-1562 \skiplinehalf \supit{b}Observatories of the Carnegie
Institution of Washington, 813 Santa Barbara St., Pasadena, CA 91101}

\authorinfo{Further author information: (Send correspondence to
  R. Michael Rich)\\
  E-mail: rmr@astro.ucla.edu $^1$Based in large part on observations obtained at the W.M. Keck Observatory, which is operated jointly by the California
Institute of Technology and the University of California}

\begin{document}

\maketitle

  \begin{abstract}
We report on detailed abundances of giants in the
Galactic bulge, measured with the HIRES echelle spectrograph
on the 10-m Keck telescope.  We also review other work on
the bulge field population and globular clusters
using Keck/HIRES.  Our new spectra have 3 times the
resolution and higher S/N than previous spectra obtained with 4m telescopes.
We are able to derive $\log g$ from Fe II lines and excitation 
temperature from
Fe I lines, and do not rely on photometric estimates for these
parameters.
We confirm that the iron abundance range extends from $-1.6$
to $+0.55$ dex.   The improved resolution and S/N of the
Keck spectra give [Fe/H] typically 0.1 to 0.2 dex higher than previous
studies,\cite{mr94} for bulge 
stars more metal rich than the Sun.
Alpha elements are enhanced even for stars
at the Solar metallicity (as is the case for bulge globular
clusters).  We confirm our earlier abundance analysis 
of bulge giants\cite{mr94} and find that Mg and Ti
are enhanced relative to Ca and Si even up to [Fe/H]=+0.55.  
We also report the first
reliable estimates of the bulge oxygen abundance.  Our element
ratios confirm  that bulge giants have a clearly identifiable
chemical signature, and
suggest a rapid formation timescale for the bulge.


  \end{abstract}

\section{Introduction}

Because of their faintness, reddening, severe crowding,
and high metallicity, the
stars of the Galactic bulge remained among the last Galactic
population to be studied with high resolution spectroscopy.
In the scientific cases for large telescopes, the goal of 
successfully defining the abundances and chemistry of bulge
stars has often figured prominently.  Of course, the real
driver for studying these stars is not the technical challenge,
rather it is their potential to yield insights into the
formation of bulges and ellipticals.

Within the last five years, the combination of spectroscopy
with the Keck telescopes
and imaging with the Hubble Space Telescope has revolutionized
the study of galaxies at high redshift.  A population of plausible
progenitors \cite{steidel96} to present-day $L^*$ galaxies has been discovered
at $z>3$ and a proposed star formation history \cite{madau96}
of the Universe has been sketched out.  
However, these observations cannot trace the evolution of the
$z>3$ galaxies into their present-day counterparts.  In many
respects, such as luminosity and clustering, they strongly
resemble the progenitors of present-day luminous galaxies.
It is also possible to constrain the formation time of bulges
from observations of galaxies at $z\leq 1$.  Recent pixel-by-pixel
analysis \cite{ellis2000} of resolved images of high redshift galaxies with
clearly visible bulges apparently shows
that at any given redshift, bulges are bluer than the
reddest galaxies of elliptical morphology.  Unfortunately, this imagery
cannot easily distinguish between a late starburst on top of an
old population versus a mostly intermediate-age population.
So it is valuable to seek other available evidence, such as
the ages and abundances of stars in the Galactic bulge.

The exact agreement between HST luminosity functions of old
metal rich globular clusters, and NTT luminosity functions of the
Galactic bulge field\cite{sergio95} strongly suggests that the
bulge formed early and rapidly.  HST photometry in a
number of different bulge fields also shows that the stars brighter
than the oldest turnoff point are foreground stars associated with
the disk, not the bulge\cite{feltz00}. Age constraints from luminosity
functions or the luminosity of the main sequence turnoff point, 
while powerful, are only accurate to (at best) $\approx 1-2 Gyr$.
The detailed composition of stars in the bulge does not constrain
the absolute age of the bulge.  However, it does constrain the 
timescale for chemical enrichment, and it helps to relate the bulge
(or not) to elliptical galaxies.  As a larger sample of stars is
accumulated, more detailed theoretical inferences about the enrichment
history will be possible.

The bulge of the Milky Way is a clearly distinct population,
as defined by the classical characteristics of a stellar population,
age, abundance, kinematics, and structure.
The central 1000 pc of our galaxy is dominated by old, metal rich
\cite{rich88}$^,$\cite{mr94}
stars with very high phase space density. 
The stellar mass of the bulge is 
$2\times 10^{10}M_\odot$, roughly 1/3 that of the
disk, but it still accounts for a large fraction of the
baryonic mass of the Galaxy. The image of the bulge\cite{hauser90} 
obtained using the DIRBE instrument on board the
COBE satellite dramatically illustrates
its distinct nature and its similarity to more distant ellipticals.
It is possible to develop a model\cite{zhao96} that both fits the
surface brightness in the COBE image, solves Poisson's equation, and gives stellar orbits
that reproduce the observed kinematics of the bulge.

Presently, there is no clear consensus on the ages and formation timescales
of bulges in general.  The colors of bulges imaged in detail in the optical
and IR by HST are consistent with very large ages\cite{pel99}, a result first
found in 1969 for the bulge of M31\cite{sandage69}.  On the other hand,
the integrated Mg line strengths of bulges are less than those of
ellipticals at the same iron line strengths\cite{proc00}, which would
argue that bulges might have experienced a less intense and more extended
period of star formation than the ellipticals.

\subsection{How Element Ratios May Constrain the Formation of the Bulge}

The motivation for measuring abundance ratios in old stars is that they
preserve the fossil record of the early star formation process.
Potentially, the initial mass function, star formation rate, and
importance of infall or extended star formation at late times can
all be recovered from abundance ratios.   The material treated briefly
below is discussed in more detail elsewhere.
\cite{andy97}$^,$\cite{mr99}  Scenarios for forming the bulge
predict a wide range of timescales, from $\sim 10^8$ yr for a violent
starburst, to a few Gyr for a massive disk that thickens into a bar.
The modeling of observed abundance trends can distinguish among these
models.

\begin{figure}[ht]
  \begin{center}\leavevmode
    \psfig{figure={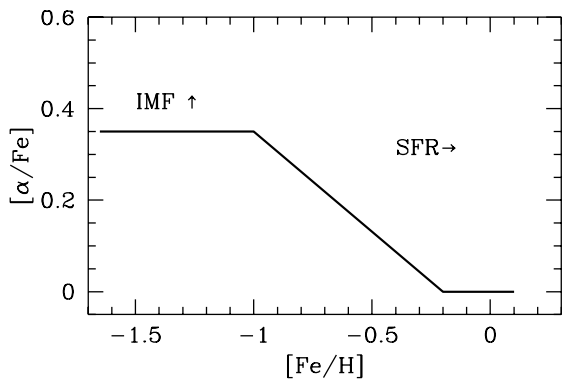},width=100mm,angle=0}
\vspace*{-5.0cm}
    \caption{A schematic plot of $\rm [\alpha/Fe]$ vs [Fe/H] in disk/halo
stars.  A shallower IMF slope (increasing number of massive stars)
will enhance the alpha elements.  A higher star formation (faster 
enrichment) will keep the alpha elements high, even above the 
Solar iron abundance.}
    \label{fig:sfr}
  \end{center}
\end{figure}

{\it Metallicity:}  The fundamental notion of chemical evolution is that
other than those light elements produced in the Big Bang, metals are
made in supernovae.  Because SNe explode in $\sim 10^6$ yr and distribute
their metals widely, it is possible to model the process as a simple
differential equation (the Simple Model\cite{ss72} of chemical 
evolution).  In the case of the bulge, the deep potential well and
likely violence of the early starburst satisfy the model assumptions, 
and the abundance distribution fits the Simple Model\cite{rich90}.
The yield is the ratio of the mass of metals produced to the total
mass locked up in long-lived stars.  In the Simple Model, the yield
is the mean metal abundance of the population.   The shallower the initial
mass function slope (more massive stars) the higher the yield.

{\it Alpha Elements:} When the first 200-inch echelle spectra of metal
poor stars and globular clusters were obtained\cite{wall62} it was noted
that some even-Z elements (O, Mg, Si, Ca, and Ti) were overabundant
by $\sim +0.3$ dex relative to the Solar Neighborhood.  These are the
so-called $\alpha$-elements, although their actual synthesis is
far more complicated than transmutation by successive capture of
helium nuclei in massive stars.   The widely accepted explanation for
these over abundances\cite{tinsley79}$^,$\cite{wheeler89}is that 
massive star (Type II supernovae) dominated the enrichment at early
times; models\cite{ww95} indicate that the ejecta of these SNe are very
rich in alpha elements.  
Although type I SNe produce the iron peak elements, their
contribution to the iron abundance becomes important only after
$\sim 1$Gyr, as time is required for the formation of a prior generation
of white dwarfs.

The diagnostic value of trends of $\rm [\alpha/Fe]$ vs [Fe/H] extend
beyond their use as a crude clock, as has been suggested for the
bulge\cite{matt90}.  If the IMF is dominated by massive stars, the
alpha elements can be enhanced by more than +0.3 dex\cite{ww95},
while a high star formation rate will result in stars of Solar iron
abundance having an alpha-enhanced composition, as appears to be the
case for the bulge.  Finally, although Ti is observed to be elevated
with the alpha elements, the nucleosynthesis calculations\cite{ww95}
predict a low yield of Ti in massive stars; this remains a problem.

{\it Neutron-Capture Elements:} The two dominant modes of neutron capture
also offer the potential to serve as clocks, and as a fossil record of
early star formation.  Supernovae (probably Type II) 
are believed to be the site of the
r-process\cite{wh92}, while the helium burning shells of AGB stars
are suspected as the site of s-process production, as was
shown in early calculations\cite{iben75}.  The Ba/Eu ratio is
is especially useful because it is sensitive to the r-process
fraction of heavy elements.  However, practical use of this diagnostic in the
bulge is somewhat complicated by the lack of weak Ba lines, although
La and Nd offer excellent possibilities as s-process indicators.
Depending on whether [Ba/Eu] as a function of [Fe/H]
approaches the s-process or the lower r-process value, one
can infer either a disk-like or halo-like (Type II SN ejecta
dominated composition) star formation
history.\cite{andy98}  In principle, evidence for r-process
nucleosynthesis indicates the presence of enrichment due to
Type II SNe which could be due either to a rapid burst of
star formation or a shallow IMF.
In the bulge, we hope
to use [Ba/Eu] and other heavy element diagnostics to test
the hypothesis\cite{wyse92} that the bulge formed from
gas initially enriched by the astration of the halo.
The heavy elements have tremendous potential to constrain
the enrichment timescale (and stellar masses responsible)
in great detail, especially in the difficult 1-5 Gyr regime\cite{busso99}.
The production of stable isotopes of some s-process elements
such as Rb turns out to be very sensitive to the temperature
of the helium burning shells of AGB stars.

The derivation of abundances from the equivalent widths
of the lines of heavy elements is done with caution.  Each
absorption line is split into multiple sub-components by
nuclear hyperfine splitting; failure to account for this
effect can lead to serious errors in the abundances.

Before turning to a discussion of our results, we point out that
our program would not have been possible without the
HIRES echelle spectrograph \cite{vogt94} as well as the
Keck telescopes.  Just now, in the year 2000, we are seeing
the successful first light of UVES at the VLT, and HDS at 
Subaru.  HIRES paved the way for these successful instruments,
at a time when the operational success of such an instrument
on a 10m telescope was far from guaranteed.

\section{Prior Work With Keck}

The first major effort on high resolution spectroscopy of
bulge giants was that of McWilliam \& Rich (1994)\cite{mr94}[MR94], which
represents the limit of what may be accomplished with 4m-class
telescopes.  This work showed that the mean iron abundance
in the bulge is $-0.25$ dex, not the $+0.3$ dex found from the
early low resolution spectroscopy\cite{rich88}.  As we discuss later,
the new Keck data may once again revise the abundance scale
somewhat upwards.  MR94 also found that the alpha elements
behave in a peculiar manner.  Mg and Ti are enhanced up
to Solar metallicity, while Ca and Si follow a more disk-like
enrichment trend.  Also, MR94 found several stars with enhancements
of the largely r-process element Eu, which is thought to be
produced by type~II SNe, consistent with the Mg over-abundances.
One aim of the new 10m science is to test
these findings.  

MR94 found that the neutron-capture elements Ba, Y, Zr scale
approximately with Fe in the solar ratios.  Although these
elements are also made by neutron-capture in massive stars,
the bulk of the solar composition is from low-mass AGB star
nucleosynthesis.
Rapid bulge enrichment by massive stars would likely have excluded
low-mass AGB stars from contributing s-process elements.  A
constraint on the bulge formation timescale is possible
if the detailed heavy-element abundance patterns can be used to
identify the role of low-mass AGB nucleosynthesis.

\subsection{Bulge Field Stars}

The CTIO 4m data were S/N=40 and R=17,000, while the
HIRES spectra range from R=45,000 to 60,000 for the most
metal rich stars.
The first goal after MR94 was to verify the surprising result that the
most metal rich bulge giants have [Fe/H]=+0.44.
As we mentioned earlier,
Castro et al. (1996)\cite{castro96} analyses the spectrum
of the bulge giant BW 4167 using 3 different methods: spectrum
synthesis, curve of growth, and classical equivalent width analysis
using the spectrum synthesis code MOOG\cite{sneden73}.  The analysis was hampered by the
S/N of the spectrum available, but still gave [Fe/H]=+0.47,
and that the most metal rich bulge giant has the same line strength
as the canonical metal rich disk giant, $\mu$~Leonis.
Castro et al. largely confirmed the high end of the bulge metallicity
scale found from the 4m data.  Even considering the low S/N of these
early data, the result is important as it was the first Keck
spectroscopy of a bulge giant.

Echelle spectroscopy of bulge main sequence stars at $V>20$ would
appear to be beyond the grasp of
the present generation of 8-10m telescopes.
However, these stars are occasionally magnified 
by factors of 10 or more by microlensing events, and current surveys
identify rising events with
enough regularity that one can confidently schedule observing
runs in the anticipation that amplified stars will
be available to observe.
Minniti has acquired KECK/HIRES spectroscopy of one such event in which
microlensing boost enhanced the effective diameter
of the Keck telescope to 15m \cite{min98}.  The capability
to measure a dwarf star gives the first Li abundance constraint
in the bulge; the Li abundance of
$A(Li)=2.25\pm0.25$ is slightly below that of the
Hyades ridgeline.
Although no conclusions can be drawn from
this single measurement, the technique is important for two
reasons.  First, Li is of course destroyed in the course of
stellar evolution.  While Li rich giants are known (even in the
bulge; our Keck spectra confirm the 180mA Li line found by MR94 in
BW~I-194), the source
of Li in giants is widely thought to be nuclear reactions in the envelope 
and is of course not primordial.  
Second, red giants in globular clusters
are established to undergo deep mixing, during which 
nuclear transmutation of certain alpha elements (O and Mg, but not
Si, Ca, and Ti) may occur\cite{kraft94}.  A detailed abundance
analysis of stars in NGC 6528\cite{carretta00} finds evidence for 
deep mixing, even in this cluster of approximately Solar metallicity.
While deep mixing apparently
does not appear to affect abundances of halo field giants, it is
important to obtain spectra of dwarfs in the bulge to be certain
that deep mixing is not affecting the derived abundances.

With microlensing surveys continuing, one may anticipate that the
use of the microlens boost technique will be of increasing importance.  
However,
one must obtain excellent spectra for each case, as once the microlensing
event concludes, there is no opportunity to repeat the
high-dispersion spectroscopy until the 30m - 100m telescopes
of the future become available.

\subsection{Bulge Globular Clusters}

As observations have improved, our view of the globular cluster
system toward the Galactic Center has changed.  Early studies
of the kinematics supported association of these metal rich
clusters with a disk-like system \cite{arm89}.  As larger samples
of these obscured globular clusters were studied, it
became apparent that the kinematics of these clusters more closely
resemble the bulge stars;\cite{min95} recent kinematic studies 
uphold this view\cite{cote99}.
Considering a new enlarged sample of these clusters, with distances,
\cite{barbuy98} find that their spatial distribution and
abundance distribution both follow the light of the Galactic bulge.

In contrast to the wide abundance range in the field, the bulge
globular clusters are simple stellar populations, that is, having
a very narrow range in age and abundances of their constituent stars.
Therefore, abundance
analysis of these clusters is especially valuable in 
understanding the 
more distant stellar populations in bulges and ellipticals which
presently may only be studied in their integrated light.
In these metal rich globular clusters, the stars at the tip of the
red giant branch are so blanketed by TiO that their $V$ band magnitudes
are fainter than those of the red clump.  The giant branch thus
has the form of an arc; even the $I$ band suffers some blanketing.
The same giant branch morphology is seen in 
the field population of the bulge \cite{rich98}.  Population
synthesis models of stellar populations must include the correct
giant branch.  Because many stars on first ascent have strong TiO
bands, the overall impact is to increase the TiO line strength
in stellar populations.  TiO bands are found throughout the
spectrum, including overlapping the very important
MgH feature at 5170A (the basis of the Faber $\rm Mg_2$ index.
Excessively strong TiO in the giants potentially could contribute
to a spurious measurement of enhanced Mg in elliptical galaxy
populations. 
It is therefore very important to understand
the underlying composition that gives rise to these descending
giant branches, and the simple stellar populations of metal rich
globular clusters are ideal for this.

The earliest abundance study\cite{barbuy99} of two giants in NGC 6553 relied
on spectra from the ESO 3.6m telescope.  NGC 6553 has the classic descending giant
branch arc, indicating near-solar metallicity; it lies 6kpc from the
Sun and has the same turnoff to HB magnitude difference as is found
in the extreme halo\cite{sergio95}.
The initial results\cite{barbuy99} are surprising: [Fe/H]=$-0.55$ and $\rm [\alpha/Fe]=+0.6$
for a total $Z$ of Solar or greater.  If correct, these findings would
require a revision in our view of nucleosynthesis in metal rich populations,
including elliptical galaxies.

However, a different result\cite{cohen99}$^,$\cite{carretta00} has been
found from spectra obtained using Keck/HIRES.
The greater aperture of Keck permits spectroscopy of
hotter stars on the red horizontal branch.  A team led by 
Cohen and Gratton analyze spectra of
5 RHB stars in NGC 6553
and 4 giants in NGC 6528.  They find
[Fe/H]=$-0.16$ and
[Ca/Fe]$\approx +0.3$ for NGC 6553\cite{cohen99}, and 
[Fe/H]=$-0.13$ with a similar alpha enhancement in NGC 6528
\cite{carretta00}.  In contrast to our new HIRES spectroscopy of bulge
giants which we report below, the effective temperature and
gravities are derived from photometry and the distance to the
clusters.  However, the basic checks such as trend of iron abundance
with excitation potential show that this has been a reasonable approach.
In NGC 6528, Carretta et al. find
star-to-star variations in O and Na are reminiscent
of the deep mixing effects noted in other clusters.  In particular,
O and Na are anti-correlated.  
These results show that the two bulge clusters have composition
and abundance similar to that of the bulge field stars at the
same iron abundance. 

The formation of
globular clusters is clearly observed at the present epoch
in merging galaxies\cite{whitmore99}.  
A simple analytic model\cite{fall00} in which globular clusters are tidally
limited and experience disk and bulge shocking can account for
the truncated mass distribution seen amongst the old Galactic
globular clusters at the present time.  Very likely, NGC 6528
and 6553 are survivors of the bulge's ancient globular cluster
population left over from its formation.

As the S/N of data improve, the derived abundances of metal rich
stars frequently increase because the continuum is defined more
clearly.  The Keck's aperture also enabled Cohen et al. to use
red horizontal branch stars some 700K hotter than the 4000K
cool red giants observed by Barbuy et al.  Nonetheless, more
work is called for on the bulge clusters and Keck, VLT, and
Subaru will contribute toward this effort in the next year.

\section{New Results for the Bulge K Giants from Keck/HIRES}

\begin{figure}[Ht]
  \begin{center}\leavevmode
    \psfig{figure={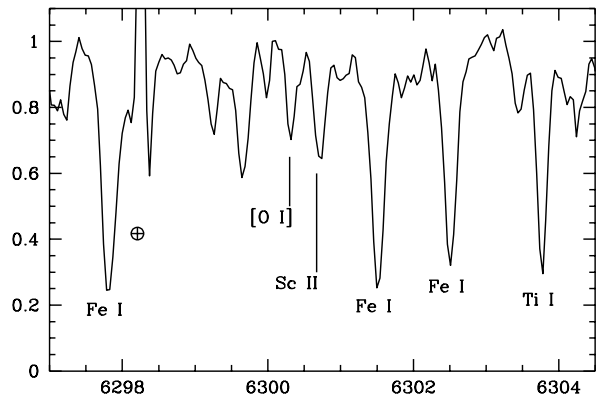},width=110mm,angle=0}
\vspace*{-4.5cm}
    \caption{This figure illustrates the spectrum of 
the most metal rich star in our sample ([Fe/H]=+0.55) BW I-039, obtained
with the HIRES spectrograph on the Keck 10m telescope.  Notice the
clean separation of OI 6300.3 from the Sc II line, which we
achieve using $R=60,000$.  This star is $V=17.5$ and was observed
through a 0.56 arcsec slit.}
    \label{fig:bw1039}
  \end{center}
\end{figure}

\begin{figure}[ht]
  \begin{center}\leavevmode
    \psfig{figure={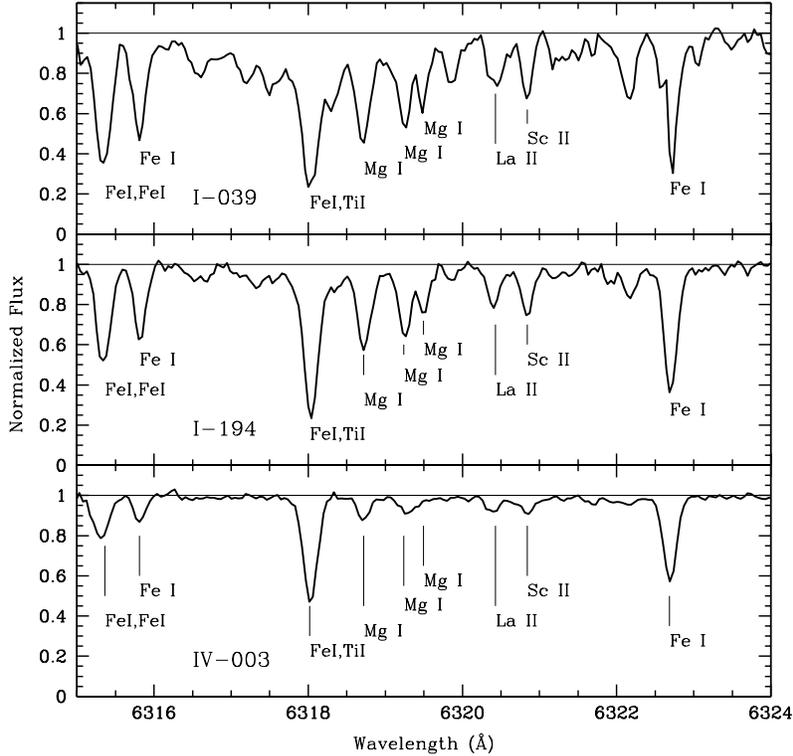},width=110mm,angle=0}
\vspace*{-0.5cm}
    \caption{Illustration of three spectra which cover
most of the abundance range found in the bulge.  
BW IV-003 has [Fe/H]=$-1.24$, while
BW I-194 has [Fe/H]=$-0.03$, and BW I-039, one of the most metal rich
stars from Rich (1988) has [Fe/H]=+0.55.  The resolution of BW I-039
is 60,000 while the other spectra are at R=45,000.  We ultimately plan
to synthesize the Mg region before deriving final abundances.}
    \label{fig:mgplot}
  \end{center}
\end{figure}

We began our Keck/HIRES spectroscopy of bulge
giants in August of 1998, with the aim
of obtaining high S/N, high resolution spectra of 25 bulge
giants with Keck/HIRES.  Initially, we are reobserving
a number of stars from our earlier study\cite{mr94}; these stars
are located in Baade's Window, a region of relatively low
extinction in the Galactic bulge some 500 pc south of the nucleus,
at $l=0^o,b=-4^o$.  At a declination of $-30^o$, the field is
accessible from Mauna Kea for about 4 hours per night.
For all but the most metal-rich stars
a 0.86 arcsec slit is used, giving $R\sim 45,000$.  For the very
metal rich stars
I$-$039 and IV$-$167 we used a 0.57 arcsec slit to obtain $R\sim 60,000$.
The data have been reduced using MAKEE, written by Tom Barlow
at Caltech.  This code has enabled us to speedily reduce these
otherwise very complicated data.  After the continuum has been
defined, the program GETJOB\cite{andy95} is run
to semi-automatically fit all measurable lines with gaussian
profiles,
to obtain equivalent widths, which are then input to
the MOOG spectrum synthesis code\cite{sneden73} using the
Kurucz \cite{kurucz92} 64 layer model atmospheres. 
Ultimately, we will synthesize small regions of spectra around 
each element of interest.  Fig. \,\ref{fig:bw1039} shows part
of the spectrum near the forbidden O~I lines, for
one of our faintest, most metal rich stars, BW I-039.  
In Fig. \,\ref{fig:mgplot}, we illustrate how the wide
abundance range present
in the bulge affects the spectra.  Spectrum synthesis of
most features will be required for stars exceeding the Solar
iron abundance.

The higher resolution and greater wavelength coverage of the HIRES
spectra offer many advantages over the MR94 study.  
In particular,
the problems of line blending and the lack of continuum regions 
is greatly improved; this is especially important for the derivation
of the oxygen abundance from the [O I] lines.  

Even so, analysis of these spectra are complicated by the
well-known problems of bulge stars.  The alpha elements,
especially Mg, are an important source of electrons in the atmospheres
of K giants.  Consequently, if [Mg/Fe]=+0.4, the 
$H^-$ continuous opacity increases relative to that
for Solar composition.  Therefore, use of simple model atmospheres with
a scaled Solar composition will give element abundances that
are spuriously low.  If bulge giants contain excess CN (and this
is very likely the case for the metal rich stars) the atmosphere
boundary temperature may be reduced enough to cause serious deviations
from the temperature structure of solar-neighborhood giants, and so
affect the abundance derived from spectrum synthesis programs which adopt solar composition model atmospheres.  One must use a grid of realistic
model atmospheres, but also one must derive as many of the stellar
parameters as possible from the spectra themselves.  

\begin{figure}[ht]
  \begin{center}\leavevmode
    \psfig{figure={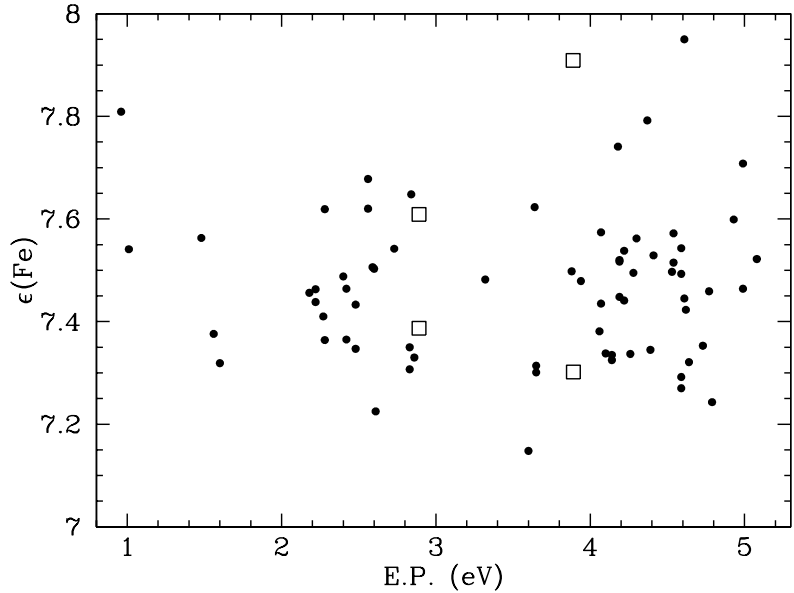},width=110mm,angle=-90}
    \caption{ Excitation plot of iron lines as a function of excitation
potential.  This plot constrains the effective temperature, abundance,
and gravity of BW I-194.  Fe I lines
are filled symbols; Fe II lines are the open squares.  The Fe II
line with the largest abundance appears to be blended. 
This star has $\rm T_{eff}=4360\pm 100K, \xi=1.23\pm 0.05\ km/sec, 
\log g=2.5, [Fe/H]=-0.03$ }
    \label{fig:exab}
  \end{center}
\end{figure}

The spectra have such good resolution and
S/N that we are able to be determine the
gravity, microturbulence, effective temperature, and [Fe/H] in a
self-consistent analysis from the Fe~I and Fe~II lines.  This is
arguably more reliable than relying upon photometric measures for
$T_{eff}$ and $\log g$ (e.g. MR94) because of the classic
problems that have plagued analysis of bulge stars:
large and spatially variable reddening, uncertainty in distance,
and at the metal rich end, blanketed broad-band colors.
Temperatures and microtubulent velocities were obtained by forcing the
iron abundance to be independent of excitation potential (Fig. \,\ref{fig:exab}) and
equivalent width, respectively; the atmosphere gravities
were adopted by requiring agreement between Fe~II and Fe~I abundances.

We find that the iron abundances are typically 0.1 to 0.2 dex
higher, an average of 0.11 dex greater, in a sample of 6 stars in
common with the MR94\cite{mr94} sample.
For example, I-194 was $-0.26$ dex,
and we find [Fe/H]=$-0.03$.  BW IV-072 was $-0.05$ dex in MR94,
but the Keck spectra give [Fe/H]=+0.25.  At the metal rich end,
BW IV-167 was found in MR94 to be +0.44 dex, and this was confirmed
in Castro et al. (1996).  We find [Fe/H]=+0.54 dex for IV-167 and
$+0.55$ dex for I-039, one of the most metal rich stars found in
the Rich (1988)
survey of 88 bulge K giants.  It is noteworthy that Castro
et al. \cite{castro96} took 3 approaches to the abundance
analysis, and that the spectrum synthesis method 
gave [Fe/H]=+0.55 for BW IV-167.
The analysis of our small sample suggests that
the mean iron abundance of the bulge may increase from the
$-0.25$ dex of MR94 to $-0.14$ dex.  The upper limit of
iron metallicity appears to be at [Fe/H]=+0.55, but obviously,
we need a larger sample of stars at the metal rich end.

The increase in iron abundance derived from the higher quality spectra
comes mainly from two sources:  Most important is that our
new continuum levels are higher, whereas MR94
could not detect the presence of weak line blanketing
(mostly from CN).  In the MR94 study, the CN blanketing
had the effect of increasing derived abundances for the weak
Fe~I lines.
This resulted in a higher microtubulent velocity for
MR94, to force stronger Fe~I lines into agreement with the weak lines;
the Keck spectra yield microtubulent velocities of 0.57 km/s lower than MR94.
The second factor is the adopted gravities: The photometric gravities of
MR94 were lower than the present spectroscopic values by an average of
$\sim 0.2$ dex, and in some cases, by as much as 0.6 dex.  The
error analysis of MR94 showed that a +0.30 dex increase in gravity
gives [Fe/H] higher by +0.05 dex.

Is this the final answer on the iron abundance?  We are beginning
to feel more confident, but we do plan to measure a large number
of weak iron lines, to confirm our findings.
Even at $R=60,000$, the continuum is not always clearly found in the 
most metal-rich stars.  We have not
yet synthesized all 8,000 CN lines (as was done in MR94) but we will
do so.  Coincidences of some Fe I lines with the occasional CN line
may bring the abundances of the most metal rich stars down slightly.
However, we believe that we are converging on the correct answer, finally.

\subsection{Relative Abundances of the Alpha Elements}

\begin{figure}[h]
  \begin{center}\leavevmode 
\vspace*{-3.0 cm} 
\psfig{figure={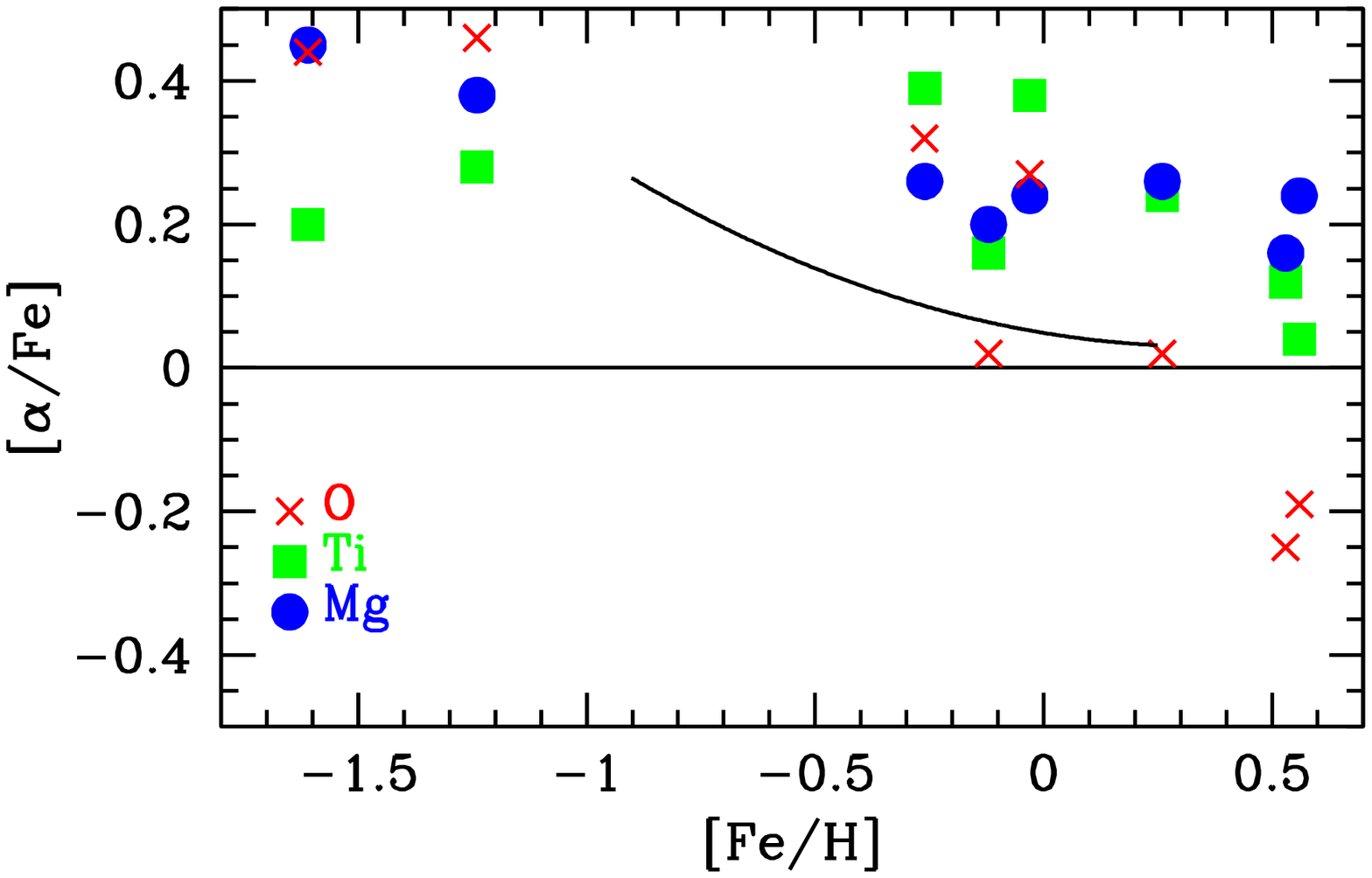},width=85mm,angle=0}
\psfig{figure={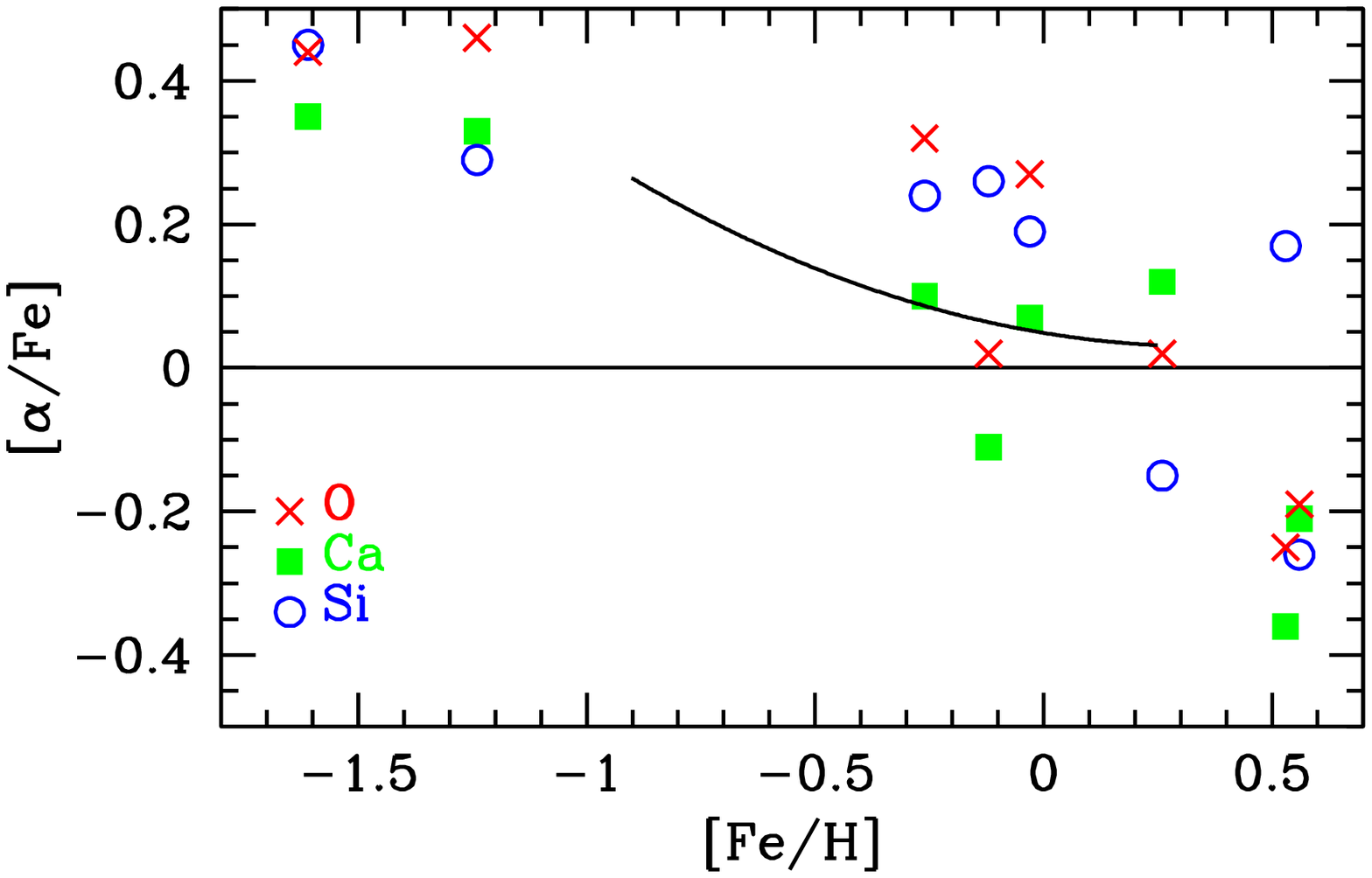},width=85mm,angle=0}
    \caption{Derived trends of element ratios for stars in the
Galactic bulge, from Keck/HIRES data.  Left panel illustrates
trends for Mg, Ti and O, while the right panel shows Ca, Si, and
O.  The solid line is the mean trend line for for all alpha elements 
in Milky Way disk stars
\cite{edvard93}. For the first time, we find a bulge giant with enhanced O
at [Fe/H]=0; a result that has been predicted\cite{matt90} but is
now seen for the first time.  Mg and Ti are clearly enhanced;
the strong enhancement of Ti is not expected from models of supernova yields
(see Fig. \,\ref{fig:yields} below).  Ca and Si follow trends that are
more typical for disk/halo stars.  Our results confirm the puzzling
abundance trends in prior studies\cite{mr94}.  Because we have not yet
done molecular equilibrium calculations, we consider the oxygen
results to be very preliminary.}
    \label{fig:ratios}
  \end{center}
\end{figure}

%See Fig.\,\ref{fig:mgti}

We now turn to the alpha elements, for which we report preliminary
abundances.  The final abundance analysis will employ spectrum
synthesis for each line region.  Returning to Figure 2,
one can inspect the Mg lines to see why this is necessary,
even for BW I-194 which has Solar abundance.
Our abundances are based generally on 2-15 lines each
of Ca, Si, and Ti, but we have only 1-3 usable Mg lines.  
%Some of the Ca and Si lines are over 150mA.  
The following results which
are based on just the equivalent width measurements should be
taken with caution.  The oxygen abundances are from the 6300.3A
forbidden line, but we consider these to be quite preliminary.
We have not yet performed the requisite CNO equilibrium calculations,
since we have not measured the carbon abundances in these stars.

However, interesting trends are beginning to emerge in 
Fig. \,\ref{fig:ratios}.  MR94
found a peculiar behavior among the alpha-elements, that Ca and Si
follow trends somewhat characteristic of the Solar neighborhood, while
Mg and Ti are enhanced as would be expected for a stellar population
enriched in a short timescale starburst.  Analysis of our first 8
stars appears to confirm MR94.  In fact, the effect appears to be even
more extreme at the metal rich end, with O joining Ca and Si.  One
interesting new result is that two stars have [O/Fe]=+0.3 at
[Fe/H]$\approx 0$.  This result was expected for the bulge\cite{matt90}, and is now
tentatively confirmed.  Disk stars\cite{edvard93} have Solar oxygen abundance at
[Fe/H]=0.

\begin{figure}[Ht]
  \begin{center}\leavevmode
    \psfig{figure={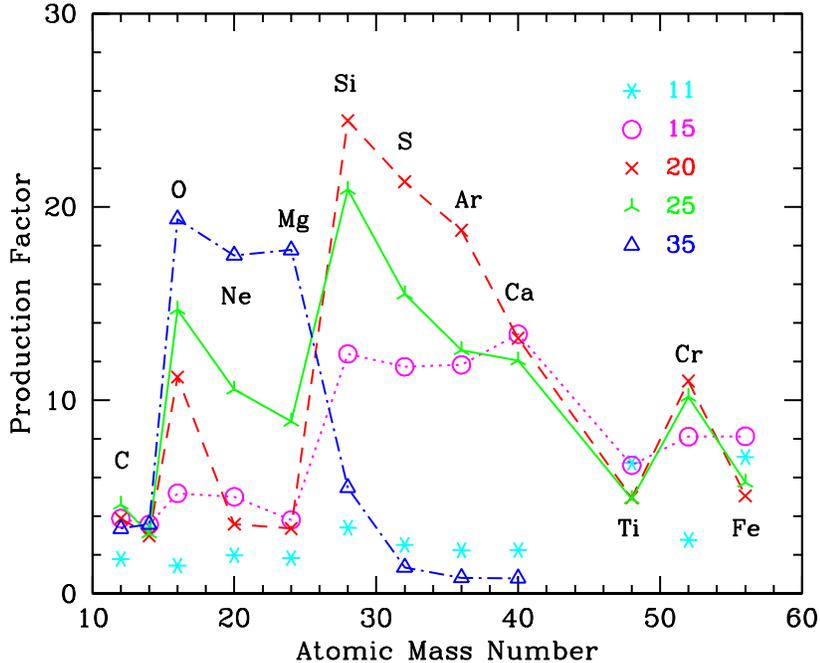},width=150mm,angle=0}
\vspace*{-4cm}
    \caption{Production factors from models of type II supernovae
by Woosley \& Weaver (1995).\cite{ww95}  
Different progenitor masses are indicated by connected symbols;
O and Mg are produced in the greatest quantities at
high mass $(\sim 35 M_\odot)$ but not in the lower mass $(15-25M_\odot)$ 
supernovae; the latter are responsible for most of the Si and
Ca production.  Note that these models predict no significant
enhancement of Ti, contrary to observations of stars in the
Galactic bulge and halo.  The production factor is the
ratio of the mass fraction of each element relative to the
total mass of the SN ejecta, divided by the element's mass
fraction in the Sun.}
    \label{fig:yields}
  \end{center}
\end{figure}

It is very premature to even speculate on the cause of the peculiar
trends among the alpha elements.  However, the source of enrichment is
supernovae, and we can turn to models of supernova yields\cite{ww95} in search of
an explanation.  The production factors in Fig. \,\ref{fig:yields} were
calculated as followed.  First, the total mass of each element produced
in the various SN models is the sum of the mass of all the stable isotopes
of that element.  Dividing the mass of each element by the mass of the
SN ejecta gives the mass fraction of that element.  The production
factor for an element is the mass fraction of that element divided
by the mass fraction of that element in the Sun.  The production
factors approximately indicate the enrichment of the ejecta relative
to Solar composition.
Fig. \,\ref{fig:yields} 
shows that O and Mg should be preferentially
produced in the most massive SNe, while Si and Ca are produced more
copiously in 15-25$M_\odot$ stars.  All of the models produce about
the same amount of Ti; presently, there are no SNe nucleosynthesis
calculations which are consistent with the observation that Ti is
enhanced in the Galactic halo and bulge populations.  Yet enhancement
of Ti is certainly real and is seen, for example, in the metal rich
globular cluster ([Fe/H=$-$0.79) M71\cite{sneden94} at the level of +0.5 dex.
The evident behavior of Ti as an alpha element remains a problem in the
modeling of supernova yields.

The incorporation of SN yields and star formation rates into
chemical evolution models gives increasingly detailed predictions
of abundance trends;
the latest of these
efforts\cite{matt99} argues for a bulge enrichment timescale of $\sim 0.5$ Gyr.
However, the physical constraint on the formation timescale is the point
at which Type I SNe begin to contribute the bulk of iron production,
which depends on the as yet unknown mechanism for Type I SNe.

\section{Summary}

In contrast to the well known achievements in the high redshift universe,
the impact of Keck on stellar abundances is less widely known, yet
significant.  Keck/HIRES spectroscopy has placed the abundance scale
of the bulge on a secure footing.  We have just begun to tap the potential
information in these spectra.  Prior efforts at measuring the oxygen
abundance in the bulge from data obtained on 4m class telescopes were
ineffective.  For the first time, we are beginning to see emerging
some clear trends in oxygen as a function of iron abundance.  The abundance 
range, and puzzling element trends found by McWilliam \& Rich (1994)\cite{mr94}
are confirmed.

Two metal rich globular clusters toward the bulge have also been the subject
of a major campaign with HIRES.\cite{cohen99}$^,$\cite{carretta00}  
NGC 6553 and
6528 have been found to have Solar metallicity with the alpha elements
of O and Ca enhanced.  The compositions of their stars are precisely
those of bulge field giants at the same metallicity.  The formation of the
proto-bulge probably proceeded much as is observed in starburst galaxies
today, with the production of numerous star clusters, a few of the more luminous of which are observed to survive to the present day.

As spectroscopy of fainter stars becomes feasible, enrichment 
trends are now available for new stellar populations, such
as dwarf spheroidal galaxies.  As more high resolution spectra
from large telescopes are analyzed, these trends may become
valuable in distinguishing the formation histories of stellar populations.
The Sagittarius dwarf spheroidal galaxy (a tidally
disrupted dwarf galaxy lying in the direction of the bulge)
is the only dwarf companion
of the Milky Way that contains stars as metal rich as the Sun.
One might speculate that the bulge could have been built from
the shards of a few such disrupted systems, and the presence
of Solar metallicity stars in the Sgr dwarf strengthens
this idea.  However, the Galactic bulge and disk populations are
dramatically different from the Sgr dwarf stars, which have
subsolar Ca and Si abundances\cite{smecker99} at [Fe/H]=0.
The trends of Mn with [Fe/H] and [Ba/Y] with [Fe/H] are
even more different between the bulge and the Sag dwarf\cite{msh00},
and it is possible to explain these differences as being
caused be early, rapid enrichment in the bulge.
The origin of the metal rich population in the Sgr
dwarf is an interesting problem in chemical evolution, given the
low mass of that galaxy and its encounter with the Milky Way.
We can pretty much rule out, however, that the metal rich population in
the Sgr dwarf was somehow captured from the bulge, or that Sgr was
once a much larger galaxy that enriched as quickly as the bulge did.

One may also compare the bulge composition to metal rich dwarf stars
in the Solar neighborhood, which are $\approx 10$ Gyr old and
reach the same high metallicities ([Fe/H]=+0.55).  High resolution
spectroscopy\cite{castro97}$^,$\cite{feltz98} 
of these stars shows them to clearly have disk-like
compositions: Mg, Ti, and O abundances are at approximately Solar
values with no clear trends.  In contrast,
the old open cluster NGC 6791 has [Fe/H]=+0.4 and enhanced Ca\cite{peterson98}.
Chemical enrichment reaching high iron abundance evidently does not
proceed the same way in all environments.  Based on the compositions
of stars, one clearly cannot produce the bulge out of the disintegrated
remnants of systems like the Sagittarius dwarf spheroidal.

Qualitatively, the abundance pattern in the bulge strongly suggests
rapid, early enrichment, consistent with the
predictions of chemical evolution models\cite{matt99}.  
The notion of rapid enrichment agrees with other studies
of the age of the stellar population\cite{sergio95} in the bulge.
The distinct nature of the bulge composition gives us confidence
that abundance ratios offer a powerful diagnostic tool that may
help to decipher the fossil record of galaxy formation.

Many open questions remain.  The bulge has a bar-like morphology,
and the most successful scenario\cite{merritt94} for forming
a bar-like bulge requires dynamical instabilities occurring in
a pre-existing disk.   However, N-body simulations of bars
indicate that they are unlikely to survive for a Hubble time,
yet the Galactic bulge is extremely old.\cite{sergio95}
Further, the extreme stellar density near the nucleus is evidence for strong dissipation being a factor in the formation of the Galactic bulge.

If the bulge abundance ratios
favor a top-heavy IMF and very rapid formation, one must infer
that ellipticals enrich more rapidly (and perhaps
with a heavier IMF) because their $\rm Mg_2$ indices at a give
$<\rm Fe>$ line strength are
so much higher compared to the bulges; in fact spiral
bulges lie near the lower range in Mg index in these diagrams
\cite{worthey92}$^,$\cite{proc00}  Before addressing these
questions, and the challenge of relating the local data to
high redshift observations, we plan to increase our sample size
and explore the behavior of different atomic species.
However, the study of the Milky Way bulge stars (and eventually, perhaps,
individual stars in the bulge of M31) does have promise
in illuminating the chemical evolution of ellipticals.

\subsection{Looking Towards the Future}

This year, two new powerful high dispersion spectrographs come on line.
At the VLT UVES has already passed science verification and has produced
beautiful data.  The HDS spectrograph at Subaru is just about to see
first light.  Fiber feeds to UVES will enable the acquisition of as
many as 8 stars in a single exposure covering all orders, or spectroscopy
of over 100 stars in a single echelle order.  The latter capability will
be enjoyed by the new echelle spectrograph that will be commissioned next
year on the Magellan I (Baade) telescope.  

On Keck, the NIRSPEC infrared spectrograph can reach $R=30,000$ in the
near-IR, and places old giants in the Galactic center within reach.
We plan to extend our abundance studies to the field and cluster stars
of the Galactic center in the next few years.

The hard reality remains that analysis of the data will still be time consuming.
For metal rich stars, it is clear that even at $R=60,000$ we require
a full spectrum synthesis before we can feel completely secure in our
results.  It will be a challenge to keep up with the flood of new
data in the coming years.  This situation should be an inspiration to
observers and theorists alike, as we enter these unprecedented times.

\section{Acknowledgments}

The entire Keck/HIRES user community owes a great debt to Jerry Nelson,
Gerry Smith, and Steve Vogt, and all of the many people have contributed
to make Keck and HIRES a reality.  We are also grateful to the W.M. Keck
foundation and to its late president, Howard Keck, for their visionary
gift that made Keck the first of a great generation of 8-10m telescopes.
We are also grateful to Tom Barlow for his MAKEE echelle extraction 
software, which made possible the rapid reduction of these data.
The work of
Andy Mcwilliam was partially supported by NSF grant AST-96-18623.
RMR acknowledges partial support from grant GO-7891 from the Space
Telescope Science Institute.  RMR is also grateful to Mark Morris,
Bob Kraft, Laura Ferrarese, and Pat Cot\'e for critical
reading of the final manuscript, and to J. Horn for
advice in formatting the manuscript.

\bibliographystyle{spiebib}

\bibliography{spi}

 \end{document}